\documentclass[technote,x11names,lettersize]{IEEEtran}
\usepackage{hyperref}
\usepackage{graphicx}
\usepackage{booktabs}
\usepackage[nocfg]{nomencl}
\makenomenclature
\usepackage{todonotes}
\usepackage{tikz}
\usepackage{gnuplot-lua-tikz} 
\usepackage{pgfplots}

\usepackage{balance}

\usepackage[utf8]{inputenc}
\usepackage{pgfplots}
\DeclareUnicodeCharacter{2212}{−}
\usepgfplotslibrary{groupplots,dateplot}
\usetikzlibrary{patterns,shapes.arrows}
\pgfplotsset{compat=newest}

\usepackage{pifont}
\newcommand{\xmark}{\ding{55}}%

\usetikzlibrary{shapes}
\usetikzlibrary{calc}
\usetikzlibrary{arrows,patterns,plotmarks,backgrounds,fit}

\usepackage{tcolorbox}
\tcbuselibrary{theorems}
\newtcbtheorem{mytheo}{Problem}%
{colback=purple!5,colframe=purple!35!black,fonttitle=\bfseries}{th}

\usepackage{xcolor}
\definecolor{lightBlue}{RGB}{166,206,227}
\definecolor{darkBlue}{RGB}{31,120,180}
\definecolor{lightGreen}{RGB}{178,223,138}
\definecolor{darkGreen}{RGB}{51,160,44}
\definecolor{lightRed}{RGB}{251,154,153}
\definecolor{darkRed}{RGB}{227,26,28}
\definecolor{lightOrange}{RGB}{253,191,111}
\definecolor{darkOrange}{RGB}{255,127,0}
\definecolor{lightViolet}{RGB}{202,178,214}
\definecolor{darkViolet}{RGB}{106,61,154}
\definecolor{yellow}{RGB}{255,255,153}
\definecolor{brown}{RGB}{177,89,40}

\definecolor{myturkish}{RGB}{141,211,199}
\definecolor{myyellow}{RGB}{255,255,179}
\definecolor{myviolet}{RGB}{190,186,218}
\definecolor{myred}{RGB}{251,128,114}
\definecolor{myblue}{RGB}{128,177,211}
\definecolor{myorange}{RGB}{253,180,98}
\definecolor{mygreen}{RGB}{179,222,105}
\definecolor{mypink}{RGB}{252,205,229}
\definecolor{mygray}{RGB}{217,217,217}
\definecolor{mypurple}{RGB}{188,128,189}
\definecolor{myseagreen}{RGB}{204,235,197}

\usepackage{glossaries}
\newacronym{3gpp}{3GPP}{Third Generation Partnership Project}
\newacronym{4g}{4G}{4\textsuperscript{th} generation}
\newacronym{5g}{5G}{5\textsuperscript{th} generation}
\newacronym{5gs}{5GS}{5G System}
\newacronym{6g}{6G}{6\textsuperscript{th} generation}
\newacronym{ai}{AI}{Artificial Intelligence}
\newacronym{a2c}{A2C}{Advantage Actor-Critic}
\newacronym{aoi}{AoI}{Age of Information}
\newacronym{api}{API}{Application Programming Interface}
\newacronym{fgan}{FGAN}{Focus Group on Autonomous Networks}
\newacronym{fr}{FR}{Frequency Range}
\newacronym{arima}{ARIMA}{AutoRegressive Integrated Moving Average}
\newacronym{bs}{BS}{Base Station}
\newacronym{bwp}{BWP}{Bandwidth Packing Problem}
\newacronym{cc}{CC}{Component Carrieer}
\newacronym{cbr}{CBR}{Constant Bitrate}
\newacronym{ccdf}{CCDF}{Complementary Cummulative Distribution Function}
\newacronym{cctc}{CCTC}{Communication-Constrained Traffic Control}
\newacronym{cots}{COTS}{Custom-Off-The-Shelf}
\newacronym{cpu}{CPU}{Central Processing Unit}
\newacronym{cv2n}{C-V2N}{Cellular Vehicle-to-Network}
\newacronym{cnn}{CNN}{Convolutional Neural Network}
\newacronym{cqi}{CQI}{Channel Quality Information}
\newacronym{ddpg}{DDPG}{Deep Deterministic Policy Gradient}
\newacronym{dl}{DL}{Downlink}
\newacronym{dz}{DZ}{Dead Zone}
\newacronym{drl}{DRL}{Deep Reinforcement Learning}
\newacronym{dqn}{DQN}{Deep Q-Network}
\newacronym{ecdf}{eCDF}{Empirical Cumulative Distribution Function}
\newacronym{elu}{ELU}{Exponential Linear Unit}
\newacronym{epdf}{ePDF}{Empirical Probability Distribution Function}
\newacronym{etsi}{ETSI}{European Telecommunications Standards Institute} \newacronym{eni}{ENI}{Experiential Network Intelligence}
\newacronym{ga}{GA}{Genetic Algorithm}
\newacronym{gfbr}{GFBR}{Guaranteed Flow Bit Rate}
\newacronym{gnb}{gNB}{gNodeB}
\newacronym{gnn}{GNN}{Graph Neural Network}
\newacronym{gps}{GPS}{Global Positioning System}
\newacronym{ilp}{ILP}{Integer Linear Programming}
\newacronym{isg}{ISG}{Industry Specification Group}
\newacronym{itu-t}{ITU-T}{International Telecommunication Union - Telecommunication standardization sector}
\newacronym{kpi}{KPI}{Key Performance Indicator}
\newacronym{lhs}{lhs}{left hand side}
\newacronym{lstm}{LSTM}{Long Short Term Memory}
\newacronym{mano}{MANO}{Management and Orchestration}
\newacronym{mcs}{MCS}{Modulation Coding Scheme}
\newacronym{mdp}{MDP}{Markov Decision Process}
\newacronym{mec}{MEC}{Multi-access Edge Computing}
\newacronym{ml}{ML}{Machine Learning}
\newacronym{mnc}{MNC}{Mobile Network Code}
\newacronym{ng}{NG}{Next Generation}
\newacronym{nfv}{NFV}{Network Function Virtualization}
\newacronym{nfvi}{NFVI}{NFV Infrastructure}
\newacronym{nn}{NN}{Neural Network}
\newacronym{nr}{NR}{New Radio}
\newacronym{nsacf}{NSACF}{Network Slice Admission Control Function}
\newacronym{pdb}{PDB}{Packet Delay Budget}
\newacronym{pi}{PI}{Proportional Integral}
\newacronym{ppo}{PPO}{Proximal Policy Optimization}
\newacronym{pdu}{PDU}{Packet Data Unit}
\newacronym{ppp}{PPP}{Poisson Point Process}
\newacronym{prb}{PRB}{Physical Resource Block}
\newacronym{qoe}{QoE}{Quality of Experience}
\newacronym{qos}{QoS}{Quality of Service}
\newacronym{ran}{RAN}{Radio Access Network}
\newacronym{rb}{RB}{Resource Block}
\newacronym{rl}{RL}{Reinforcement Learning}
\newacronym{rt}{RT}{Real-Time}
\newacronym{ru}{RU}{Radio Unit}
\newacronym{ric}{RIC}{Radio Intelligent Controller}
\newacronym{sarsa}{SARSA}{State–Action–Reward–State–Action}
\newacronym{sla}{SLA}{Service Level Agreement}
\newacronym{sdn}{SDN}{Software-Defined Networking}
\newacronym{se}{SE}{Spectral Efficiency}
\newacronym{sinr}{SINR}{Signal to Interference plus Noise Ratio}
\newacronym{smf}{SMF}{Session Management Function}
\newacronym{snr}{SNR}{Signal to Noise Ratio}
\newacronym{tes}{TES}{Triple Exponential Smoothing}
\newacronym{td}{TD}{Temporal Difference}
\newacronym{tod}{ToD}{Tele-operated Driving}
\newacronym{tovac}{TOVAC}{Tele-operated Vehicle Admission Control and Routing}
\newacronym{ul}{UL}{Uplink}
\newacronym{urllc}{URLLC}{Ultra Reliable Low Latency Constrain}
\newacronym{ue}{UE}{User Equipment}
\newacronym{vbr}{VBR}{Variable Bitrate}
\newacronym{v2x}{V2X}{Vehicle-to-Everything}
\newacronym{v2n}{V2N}{Vehicle-to-Network}
\newacronym{v2v}{V2V}{Vehicle-to-Vehicle}
\newacronym{vnf}{VNF}{Virtualized Network Function}
\newacronym{zsm}{ZSM}{Zero touch network \& Service Management}

\ifCLASSINFOpdf
\else
\fi
\usepackage{amsmath}
\usepackage{amsthm}
\usepackage{amsfonts}
\DeclareMathOperator*{\argmin}{arg\,min}

\newtheorem{corollary}{Corollary}

\newtheorem{problem}{Problem}

\usepackage{algorithm}
\usepackage{algpseudocode}
\ifCLASSOPTIONcompsoc
  \usepackage[caption=false,font=normalsize,labelfont=sf,textfont=sf]{subfig}
\else
  \usepackage[caption=false,font=footnotesize]{subfig}
\fi
\begin{document}
%
\title{\acrshort{tovac}: Tele-operated Vehicle\\Admission Control and Routing }
%
%
%

\author{Jorge Martín-Pérez,
        Carlos M. Lentisco, Luis~Bellido, Ignacio~Soto, and~David~Fernández
\thanks{This work was supported by the Remote Driver project (TSI‐065100‐2022‐003)
funded by  Spanish Ministry of Economic Affairs and Digital Transformation,
and the ECTICS project (PID2019-105257RB-C21), funded by the Spanish Ministry of Economy and Competitiveness, and the Spanish Ministry of Science and Innovation.}
\thanks{The authors are with ETSI Telecomunicaci\'on, Universidad
Polit\'ecnica de Madrid, Spain,
e-mail: jorge.martin.perez@upm.es.}
\thanks{©2023 IEEE. Personal use of this material is permitted. Permission from IEEE
must be obtained for all other uses, in any current or future media, including
reprinting/republishing this material for advertising or promotional purposes,
creating new or redistribution to servers or lists, or reuse of any copyrighted
component of this work in other work}
}

%
%

\markboth{Journal of \LaTeX\ Class Files,~Vol.~14, No.~8, August~2015}%
{Shell \MakeLowercase{\textit{et al.}}: Bare Demo of IEEEtran.cls for IEEE Journals}
%



\maketitle


\begin{abstract}

    \gls{tod} is a challenging use case
    for mobile network operators.
    Video captured by the built-in vehicle cameras must be streamed meeting a latency requirement of 5\,ms with a 99.999\% reliability.
    Although 5G offers high bandwidth, ultra-low latencies and
    high reliability; \gls{tod} service
    requirements are violated due to bad channel conditions.
    Ignoring the channel state may lead to over-estimate
    the number of \gls{tod} vehicles that can meet the service
    requirements, hence comprising the vehicle security.
    To fill this gap, 
    in this letter we propose \acrshort{tovac},
    an algorithm that guarantees \acrshort{tod} service
    requirements by taking adequate admission control
    and routing decisions. This is achieved by using
    a channel-based capacity graph that
    determines the maximum number of vehicles that can be tele-operated in any road section.
    We evaluate \acrshort{tovac} considering cellular
    deployments from Turin and show that,
    unlike a state of the art solution,
    \acrshort{tovac} guarantees the \gls{tod} service requirements.

\end{abstract}



%
\IEEEpeerreviewmaketitle

\section{Introduction}

\IEEEPARstart{T}{he} aim of \gls{tod} vehicles is to allow drivers to operate vehicles from safer and more comfortable locations, or to provide remote operator assistance in situations where an automated driving system is facing ambiguous situations. Remotely operated vehicles send their status and environment information (e.g., gathered with cameras, LIDAR and other sensors) to the remote operator through a wireless network channel. This information is presented to the operator, which sends back control or assistance commands to the vehicle through the network~\cite{amadorToD}.

The exchange of information between the remote operations centre and the tele-operated vehicles requires a network service that meets the \gls{qos} requirements defined in~\cite{3gppV2X}. \gls{qos} degradation can pose a significant threat to road safety, which translates into the need to adapt the vehicular application profile to QoS variations, for example, by decreasing the video streaming bitrate~\cite{streamingQos} and reducing vehicle speed or even making an emergency stop.


5GAA has explored how a remote driver should be warned in these situations to adapt \gls{tod} to the existing level of \gls{qos}~\cite{5GAAqos}.
However, the standards are yet far from preventing the
\gls{qos} degradation. Although the \gls{3gpp} \gls{nsacf}~\cite{3gppSystemArch}
limits the admission per network slice, it
does not prevent too many terminals from registering
simultaneously to the same cell. Thus,
the \gls{nsacf} cannot prevent \gls{qos} degradation
due to congestion.

To avoid video quality degradations, Liu \textit{et al.}~\cite{routeplan2022}
propose a \emph{static} routing and traffic control framework that decides:
the vehicle routes, the maximum distance between vehicles,
their speed, and the bandwidth  
allocated at each cell to the remote driving service.
That is, authors propose adjusting the video bitrate of each vehicle to the available bandwidth at each cell,
neglecting the experienced latency and reliability.

However, the \gls{qos} of a \gls{tod} service can also be affected by latency and reliability, as discussed by Liu \textit{et al.} in~\cite[\S III, \S VI]{routeplan2022}.
Motivated by such statement, this letter describes \gls{tovac}, an admission control and routing algorithm that also meets
the latency and reliability requirements defined by the \gls{3gpp} for \gls{tod}~\cite[Table 5.5-1]{3gppV2X}: ($i$) a~\gls{pdb} shorter than 5\,ms; and ($ii$) a
99.999\% reliability.
Additionally, \gls{tovac} ($iii$) does not impose a speed reduction of
vehicles; and ($iv$) solves the admission control problem
avoiding overlaps in time.

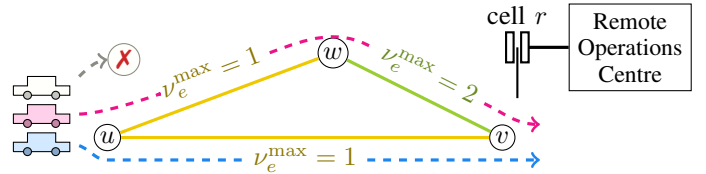
\begin{figure}[t]
    \centering
    \begin{tikzpicture}
    \def\getangle(#1)(#2)#3{%
      \begingroup%
        \pgftransformreset%
        \pgfmathanglebetweenpoints{\pgfpointanchor{#1}{center}}{\pgfpointanchor{#2}{center}}%
        \expandafter\xdef\csname angle#3\endcsname{\pgfmathresult}%
      \endgroup%
    }

    \def\carw{.75};
    \def\carh{.25};
    \def\carx{0};
    \def\cary{0};
    \path[draw,fill=DodgerBlue1!20] (\carx,\cary) -- ($(\carx+\carw,\cary)$) --
        ($(\carx+\carw,\cary+\carh/2)$) node (morrod) {} --
        ($(\carx+\carw-\carw/4,\cary+\carh/2)$) --
        ($(\carx+\carw-\carw/4,\cary+\carh)$) node (frontcar) {} --
        ($(\carx+\carw/4,\cary+\carh)$) --
        ($(\carx+\carw/4,\cary+\carh/2)$) --
        ($(\carx,\cary+\carh/2)$) --
        (\carx, \cary)
        ;
    \filldraw[fill=DodgerBlue4!50] ($(\carx+\carw/4,\cary)$) circle (1.5pt);
    \filldraw[fill=DodgerBlue4!50] ($(\carx+\carw*3/4,\cary)$) circle (1.5pt);

    \def\caruw{.75};
    \def\caruh{.25};
    \def\carux{0};
    \def\caruy{1.5*\carh};
    \path[draw,fill=DeepPink1!20] (\carux,\caruy) -- ($(\carux+\caruw,\caruy)$) --
        ($(\carux+\caruw,\caruy+\caruh/2)$) node (morrou) {} --
        ($(\carux+\caruw-\caruw/4,\caruy+\caruh/2)$) --
        ($(\carux+\caruw-\caruw/4,\caruy+\caruh)$) node (frontcaru) {} --
        ($(\carux+\caruw/4,\caruy+\caruh)$) --
        ($(\carux+\caruw/4,\caruy+\caruh/2)$) --
        ($(\carux,\caruy+\caruh/2)$) --
        (\carux, \caruy)
        ;
    \filldraw[fill=DeepPink4!50] ($(\carux+\caruw/4,\caruy)$) circle (1.5pt);
    \filldraw[fill=DeepPink4!50] ($(\carux+\caruw*3/4,\caruy)$) circle (1.5pt);

    \def\caruuw{.75};
    \def\caruuh{.25};
    \def\caruux{0};
    \def\caruuy{3*\carh};
    \path[draw,fill=Ivory1!20] (\caruux,\caruuy) -- ($(\caruux+\caruuw,\caruuy)$) --
        ($(\caruux+\caruuw,\caruuy+\caruuh/2)$) node (morro) {} --
        ($(\caruux+\caruuw-\caruuw/4,\caruuy+\caruuh/2)$) --
        ($(\caruux+\caruuw-\caruuw/4,\caruuy+\caruuh)$) node (frontcaruu) {} --
        ($(\caruux+\caruuw/4,\caruuy+\caruuh)$) --
        ($(\caruux+\caruuw/4,\caruuy+\caruuh/2)$) --
        ($(\caruux,\caruuy+\caruuh/2)$) --
        (\caruux, \caruuy)
        ;
    \filldraw[fill=Ivory4!50] ($(\caruux+\caruuw/4,\caruuy)$) circle (1.5pt);
    \filldraw[fill=Ivory4!50] ($(\caruux+\caruuw*3/4,\caruuy)$) circle (1.5pt);

    \node[draw,circle,inner sep=1pt] (u)
        at ($(\carx+\carw+.5,\cary*.5+\caruy*.5)$) {$u$};
    \node[draw,circle,inner sep=1pt] (w)
        at ($(u)+(3,1.5*\caruuy)$) {$w$};
    \node[draw,circle,inner sep=1pt] (v)
        at ($(w |- u)+(2.25,0)$) {$v$};

    \getangle(u)(v)A;
    \getangle(u)(w)B;
    \getangle(w)(v)C;
    \draw[Gold2,very thick] (u)
        -- node[pos=.5,rotate=\angleA,anchor=north,
        Gold4] (numaxuv) {$\nu_e^{\max}=1$} (v);
    \draw[Gold2,very thick] (u)
        -- node[pos=.5,rotate=\angleB,anchor=south,
        Gold4] (numaxuw) {$\nu_e^{\max}=1$} (w);
    \draw[OliveDrab3,very thick] (w)
        -- node[pos=.5,rotate=\angleC,anchor=south,
        OliveDrab4] (numaxwv) {$\nu_e^{\max}=2$} (v);



    \def\ruh{.8};
    \node (ruBase) at ($(v)+(.2,.4)$) {};
    \draw[thick] (ruBase) -- ++(0,\ruh)
        node (ruTop) {}
        node[pos=1.25,anchor=south,align=center] {cell $r$};
    \draw[thick] ($(ruBase)+(-.15,\ruh*3/4)$) rectangle ++(.1,.4)
        node (panelleft) {};
    \draw[thick] ($(ruBase)+(.05,\ruh*3/4)$) rectangle ++(.1,.4)
         node (panelright) {};


    \node[align=center,draw,font=\small] (app)
        at ($(ruTop)+(1.5,0)$)
        {Remote\\Operations\\Centre} ;
    \draw[very thick]
        (panelright |- app) -- (app);

    \draw[DodgerBlue2,dashed,very thick]
        (morrod) to[out=-20,in=180]
        (u.south |- numaxuv.west) -- (numaxuv.west);
    \draw[DodgerBlue2,dashed,very thick,->] (numaxuv.east) --
        ($(v.south |- numaxuv.east)+(.5,0)$);
    \draw[DeepPink2,dashed,very thick]
        (morrou) to[out=0,in=\angleB] (numaxuw.west);
    \draw[DeepPink2,dashed,very thick]
        (numaxuw.east) to[out=\angleB,in=150] (numaxwv.west);
    \draw[DeepPink2,dashed,very thick,->] (numaxwv.east)
        to [out=\angleC,in=180]
        ($(v.north)+(.5,0)$);


    \draw[->,dashed,very thick,Ivory4] (morro) to[out=45,in=190] 
        ++($(.5,\carh+.1)$) node[anchor=west,circle,
            fill=Ivory1!20,solid,draw,very thick,
        inner sep=1pt, draw,thin,text=Firebrick3] {\xmark};

\end{tikzpicture}
    \caption{
    Illustration of the
    edges within
    \gls{tovac} capacity
    graph, each
    with the maximum number
    of \gls{tod} vehicles
    admitted $\nu_e^{\max}$.
    Further roads have smaller
    capacity (yellow) due
    to path loss, while a road
    nearby the cell~$r$ (green)
    has higher capacity.
    Consequently,
    TOVAC just admits the first two vehicles and rejects the third (top vehicle),
    since cell~$r$ only guarantees
    the \gls{tod} service requirements for a single
    vehicle $\nu_e^{\max}=1$ at roads $(u,w),(u,v)$.}
    \label{fig:intro}
\end{figure}

\gls{tovac} calculates the channel capacity using the 99.999-percentile of the \gls{sinr} to determine the maximum
number of vehicles that can be tele-operated along the roads while meeting the 5\,ms \gls{pdb} requirement.
Specifically, \gls{tovac} creates a capacity graph such as the one illustrated in Fig.~\ref{fig:intro}, with each
edge $e$ representing a road with an associated capacity of $\nu_e^{\max}$ tele-operated vehicles.
\gls{tovac} receives, concurrently, several requests to obtain their routes (three in Fig.~\ref{fig:intro}), finding the shortest path for each vehicle without incurring into \gls{qos} violations. As Fig.~\ref{fig:intro} shows, \gls{tovac} admits the first two vehicles but rejects the third (the white vehicle at the top) because the serving cell $r$
    does not have enough capacity $\nu_e^{\max}$ to meet the \gls{qos} requirements at roads $(u,w)$ and $(u,v)$.



\section{\gls{nr} \gls{tod} capacity model}
\label{sec:model}

In this section we describe a model to obtain the maximum number
of \gls{tod}
vehicles at each road segment. First, we obtain the radio
resources needed to meet the service \gls{pdb} at a road segment. Then, we
compute the 99.999\% spectral efficiency for a vehicle at a specific coordinate served by a cell.
Lastly, we use the two previous results to create
a road capacity graph whose weights
are the maximum number of vehicles that have enough radio resources to meet the \gls{pdb}
with 99.999\% probability.

\subsection{Radio resources for remote driving}
\label{subsec:resources}

At the radio resources level, our approach is to guarantee the availability of enough \gls{rb}s to send the \gls{tod} service packets within the timespan of a \gls{pdb}. Assuming a packet size of $L$ bits, each packet consumes:

\begin{equation}
    R(S_r^\nu,\mu,L) =
    \left\lceil
        \frac{L}{T_S^\mu\ 12 \Delta f^\mu
        \ S_r^\nu}
    \right\rceil
    \text{[RB/packet]}
    \label{eq:prb-per-udp}
\end{equation}
with $T_S^\mu=\tfrac{10^{-3}}{14\cdot2^\mu}$~[sec] symbol
duration with numerology $\mu$,
$12\cdot\Delta f^\mu$~[Hz/RB] the bandwidth of an \gls{rb}
using 12 subcarriers in the frequency domain,
and $S_r^\nu$ the \gls{se}~[bps/Hz] for a vehicle $\nu$ that is served by cell $r$.

Considering a maximum service bitrate of $b$~[bps] during any timespan of $t$~[sec],
each vehicle $\nu$ will send 
$\left\lceil \tfrac{b \cdot t}{L}\right\rceil$~packets of
size $L$.
Hence, in $t$~sec each vehicle $\nu$ generates:
\begin{equation}
    \alpha_\nu(t,S_r^\nu,\mu,b, L)=R(S_r^\nu,\mu,L)
    \left\lceil\tfrac{b \cdot t}{L}
    \right\rceil~\text{[RB]}
    \label{eq:vehicle_rb}
\end{equation}



For a set of vehicles $N_r$ requesting service from cell~$r$ we can select the maximum number of \gls{rb} generated by any vehicle to obtain the maximum number of vehicles that can be served meeting a \gls{pdb} of $D$:

\begin{equation}
    \nu_r^{max}(D,\mu,b,L)=\left\lfloor \frac{\lfloor (1-OH)\cdot\text{NRB}_r(D,B) \rfloor}{ \max_{\nu \in N_r} \alpha_\nu(D,S_r^\nu,\mu,b,L)} \right\rfloor~\text{[veh]}
    \label{eq:nu-max}
\end{equation}
In~\eqref{eq:nu-max}
$OH$ is the signaling overhead in \gls{nr} and
$\text{NRB}_r(D,B)$ the number of \gls{rb}s available for the \gls{tod} service  in $D$~[sec] at a cell $r$ using a bandwidth of $B$~[MHz].


\subsection{Cell \acrlong{se}}
\label{subsec:se}

With~\eqref{eq:nu-max} we obtain a worst-case value for the number of vehicles that can be served by a  cell. But note that the number of \gls{rb}s that each vehicle
needs depend on the \gls{se}~\eqref{eq:prb-per-udp}, with lower \gls{se} leading to a larger number of required~\gls{rb}s.

To obtain~\gls{se} we first need to compute the \gls{sinr}.
Specifically, we obtain the \gls{sinr} \gls{ccdf} to 
know, e.g., whether a vehicle experiences a \gls{sinr}
$\gamma_{r,\nu}=20$~dB with 99.999\% probability.
We resort to a probabilistic model as the proposed in
\cite{snc,coverage} to obtain the following expression.

\begin{corollary}
    If the cell power follows an exponential
    distribution $h\sim exp(\mu)$,
    and the interfering cells power is
    also exponential and identically distributed
    $g_i\underset{i.i.d}{\sim} exp(\lambda)$;
    the \gls{sinr} \gls{ccdf} satisfies
    \begin{equation}
        \mathbb{P}\left[
        \gamma_{r,\nu}>\hat{\gamma}\right]=
        e^{-\mu\hat{\gamma}d_{r,\nu}^{\alpha}\sigma^2}
        \prod_i \frac{\lambda}{\lambda+\mu \hat{\gamma} d_{r,\nu}^\alpha d_{r_i,\nu}^{-\alpha}}
        \label{eq:ccdf}
    \end{equation}
with $\alpha>0$ the path loss exponent, $\sigma^2$ the value of additive and constant noise power,
and $d_{r,\nu}$ the distance between cell $r$ and vehicle
$\nu$.
    \label{corol:snr}
\end{corollary}
\begin{proof}
    The \gls{sinr} is expressed as
    $\gamma=h d_{r,\nu}^{-\alpha}(\sigma^2+I_r)^{-1}$ with
    $hd_{r,\nu}^{-\alpha}$ the received power,
    $h\sim exp(\mu)$, and
    $I_{\nu}=\sum_i g_i d_{r_i,\nu}^{-\alpha}$ being the
    neighboring interference experienced by the 
    vehicle $\nu$ assuming $g_i\underset{i.i.d}{\sim} exp(\lambda)$.
    Hence
    \begin{multline}
        \mathbb{P}(\gamma>\hat{\gamma})
        =\mathbb{P}(h>\hat{\gamma} d_{r,\nu}^\alpha (\sigma^2+I_{\nu}))\\
        =\mathbb{E}_{I_{\nu}}\left[ \mathbb{P}(h>\hat{\gamma} d_{r,\nu}^\alpha (\sigma^2+I_{\nu})|\ I_{\nu}) \right]
        =\mathbb{E}_{I_{\nu}}\left[ e^{-\mu\hat{\gamma}d_{r,\nu}^\alpha(\sigma^2+I_{\nu})} \right]\\
        =e^{-\mu\hat{\gamma}d_{r,\nu}^\alpha\sigma^2}\cdot\mathcal{L}_{I_{\nu}}(\mu\hat{\gamma}d_{r,\nu}^{\alpha})
        \label{eq:ccdf-proof}
    \end{multline}
    with $\mathcal{L}_{I_{\nu}}(s)$ the Laplace transform of the
    neighboring interference, which has the following expression
    \begin{multline}
        \mathcal{L}_{I_{\nu}}(s)
        =\mathbb{E}_{I_{\nu}}[e^{-s I_{\nu}}]
        =\mathbb{E}_{\{g_i\}}\left[e^{-s \sum_i g_i d_{r_i,\nu}^{-\alpha}}\right]\\
        =\mathbb{E}_{\{g_i\}}\left[\prod_i e^{-s g_i d_{r_i,\nu}^{-\alpha}}\right]
        \underset{\substack{\{g_i\}\\i.i.d}}{=}\prod_i \mathbb{E}_{g_i}\left[e^{-s g_i d_{r_i,\nu}^{-\alpha}}\right]\\
        =\prod_i \int_0^\infty e^{-s g d_{r_i,\nu}^{-\alpha}}
        \lambda e^{-\lambda g}\ dg
        = \prod_i \frac{\lambda}{\lambda + s\cdot d_{r_i,\nu}^{-\alpha}}
        \label{eq:laplace}
    \end{multline}
    By plugging \eqref{eq:laplace} into \eqref{eq:ccdf-proof},
    \eqref{eq:ccdf} holds.
\end{proof}

Using Corollary~\ref{corol:snr}
we compute the 99.999\%
\gls{sinr} as $\hat{\gamma}: \mathbb{P}(\gamma\geq\hat{\gamma})\geq0.99999$.
To compute the 99.999 percentile \gls{se} $S_r(\hat{\gamma})$
we use the Shannon capacity expression. 

Given the 99.999 percentile \gls{se}, we just have to check
which \gls{nr} \gls{mcs} provides the closest \gls{se}.
Specifically, we obtain the 99.999 percentile effective \gls{se}
as:
\begin{equation}
    \hat{S}_r(\hat{\gamma}) = 
    \argmin_{s\in S:\ S_r(\hat{\gamma})>s}
    \left\{ S_r(\hat{\gamma})-s \right\}
    \label{eq:nr-se}
\end{equation}
with $s$ the \gls{se} achieved by each \gls{mcs}
reported in~\cite[Table~6.1.4.1]{3gpp-modulation}.

\subsection{Road capacity graph}
\label{subsec:graph}

With Sections~\ref{subsec:resources} and
\ref{subsec:se} we now calculate the maximum number of
\gls{tod} vehicles that can be operated at each road.
Specifically, we are interested in obtaining a weighted graph $G$ with
the roads as edges $e\in E(G)$ and the the maximum number of vehicles that can be operated at each road $\nu_e^{\max}$ as weights.

We define $E(r)\subset E(G)$ as the set of
roads covered by cell $r$. For each road $e$ served by cell $r$ we select the worst effective \gls{se}:
\begin{equation}
    \hat{S}_{r,e}(\hat{\gamma})=
    \min_{x\in e} \hat{S}_r(\hat{\gamma}_x)
    \label{eq:road-SE}
\end{equation}
with $\hat{\gamma}_x$
the 99.999 percentile \gls{sinr} provided by cell $r$
at coordinate $x$ within road $e$.

Using the the worst effective \gls{se} calculated with~\eqref{eq:road-SE} we can re-formulate~\eqref{eq:nu-max} to obtain the maximum number of vehicles served by cell $r$ at road $e$:
\begin{equation}
    \nu_{r,e}^{max}(D,\mu,b,L)=\left\lfloor \frac{\lfloor \iota_r(e)\cdot(1-OH)\cdot\text{NRB}_r(D,B) \rfloor}{ \alpha_\nu(D,\hat{S}_{r,e}(\hat{\gamma}),\mu,b,L)} \right\rfloor
    \label{eq:max-road-vehicles-gnb}
\end{equation}
with $\iota_r(e)\in[0,1]$ being the portion of \gls{rb}s that
cell $r$ gives to road $e$.

Hence, the total maximum
number of vehicles that can be operated at each road is
\begin{equation}
    \nu_e^{\max}(D,\mu,b,L)=
    \!\!\!\!\!
    \sum_{r: e\in E(r)}\!\!\!\!
    \nu_{r,e}^{max}(D,\mu,b,L)
    \label{eq:max-road-vehicles}
\end{equation}
The purpose of the capacity graph $G$ is to use it for
the admission control and routing of \gls{tod} vehicles.
All we have to ensure is that there are no more than
$\nu_e^{\max}$ vehicles at each road $e$.

\section{\gls{tod} Admission Control and Routing}
\label{sec:urac}

In the previous section, the maximum number of \gls{tod} vehicles that can be operated on each road section has been determined. But there is still a need for a mechanism to control the path that vehicles follow along the road, so that the \gls{tod} service is denied if a vehicle attempts to enter a road section that has reached its maximum capacity.

We model this problem as a multi-commodity flow problem over time~\cite{multi-commodity-over-time}
on the capacity graph $G$ defined in
Section~\ref{subsec:graph}.

\begin{problem}[\gls{tod} Admission Control and Routing]
    Given a fleet of \gls{tod} vehicles
    with their sources
    $\{s_\nu\}_\nu\subset G$ and destinations
    $\{d_\nu\}_\nu\subset G$, solve:
    \begin{align}
        \max_{X_{\mathcal{P}_\nu}^\nu} &
        \sum_{\nu,\mathcal{P}_\nu}
        X_{\mathcal{P}_\nu}^\nu\\
        s.t.: & 
        \sum_{\nu,\mathcal{P}_\nu}
        X_{\mathcal{P}_\nu}^\nu(e,t)\leq \nu_e^{\max}(D,\mu,b,L),\quad \forall t>0,e\in G\label{eq:capacity}
    \end{align}
    with $X_{\mathcal{P}}^\nu(e,t)\in\{0,1\}$
    an auxiliary variable taking 1 if
    $e\in\mathcal{P}\land
    t\in[\tau_\nu(e^+),\tau_\nu(e^-))$;
    and $\mathcal{P}_\nu$ a path with
    origin $s_\nu\in G$ and destination $d_\nu\in G$.
    \label{problem}
\end{problem}

By solving the Problem~\ref{problem}, the number of vehicles flowing from their sources $\{s_{\nu}\}_\nu$ to their destinations $\{d_{\nu}\}_\nu$ is maximized.  Constrain \eqref{eq:capacity} ensures that no section of the road ($e$) carries more vehicles than those set by the capacity graph $\nu_e^{\max}(D,\mu,b,L)$.      
Therefore, constraint \eqref{eq:capacity} ensures
the \gls{tod} latency and reliability requirements.

In the formulation of the problem, each path is denoted as $\mathcal{P}$. The binary decision variable $X_\mathcal{P}^\nu(e,t)$ takes the value one if the vehicle $\nu$ is traversing the road section $e$ at instant $t$, where $e$ belongs to the path $\mathcal{P}$ followed by the vehicle. The travel time of a vehicle in a road section is calculated as $\tau(e)=\tau_\nu(e^-)-\tau_\nu(e^+)$, being $\tau_\nu(e^+)$ and $\tau_\nu(e^-)$ the times at which the vehicle enters/exits road $e$.
Hence, $\tau_\nu(e^+),\tau_\nu(e^-)$ are computed as follows:
\begin{align}
    \tau_\nu(e^+)=&\sum_\mathcal{P}\sum_{e'\in\mathcal{P}: e'\prec e}X_{\mathcal{P}}^\nu\tau(e')\\
    \tau_\nu(e^-)=&\tau_\nu(e^+)+\tau(e)
\end{align}
with $e'\prec e$ denoting that road $e'$
preceeds road $e$ in path $\mathcal{P}$.

Therefore, the total travel time of a vehicle, denoted as  $\tau_\nu(\mathcal{P})$, can be calculated as the sum of the travel times on the road sections belonging to the path, i.e., as $\tau_\nu(\mathcal{P}) = \sum_{e\in\mathcal{P}}\tau(e)$.

To solve Problem~\ref{problem} we propose Algorithm~\ref{al:astar} (\gls{tovac}), a modified version of the $A^*$ search algorithm. The $A^*$ algorithm has already been used in the context of \gls{tod}~\cite{routeplan2022} and fits well in network graphs composed by a large amount of nodes~\cite{astar1}. This is mainly because the search is guided by a heuristic function (the Haversine distance is the most commonly used function). To further reduce the time taken by the algorithm to calculate the shortest path in our scenario, all the roads not meeting the service requirements are pruned from the network graph (line 1 in Algorithm~\ref{al:astar}).

By using the $A^*$ algorithm, the shortest path for each vehicle $\nu$ from a source $s_\nu$ to a destination $d_\nu$ is determined (line 3).
We have modified the $A^*$ algorithm so that it satisfies constraint~\eqref{eq:capacity}.
The idea is that by using a weight function $\omega_\nu$, defined in Eq.\eqref{eq:distance}, the $A^*$ algorithm will
not steer vehicles through routes that contain road sections that cannot admit more vehicles. Instead, the $A^*$ algorithm will choose paths having the radio resources needed to support the vehicle's communications while meeting the latency requirement of the service ($D$\,ms) with a reliability level of $99.999\%$.  

The weigh function defined in Eq.\eqref{eq:distance} assigns a road section a cost of infinity if its capacity ($\nu_e^{\max}(D,\mu,b,L)$) is exhausted
\begin{align}
    \omega_\nu(e)=\begin{cases}
        \tau(e), & V_\nu(e)<\nu_e^{\max}(D,\mu,b,L)\\
        \infty, & V_\nu(e)\geq\nu_e^{\max}(D,\mu,b,L)
    \end{cases}
    \label{eq:distance}
\end{align}
with $V_\nu(e)$ denoting the number of vehicles that
traverse road section $e$ at the same time as vehicle
$\nu$:
\begin{equation}
V_\nu(e)=\left|\nu'\neq\nu: [\tau_{\nu'}(e^+),\tau_{\nu'}(e^-))
        \cap [\tau_\nu(e^+),\tau_\nu(e^-)) \neq\emptyset\right|
    \label{eq:overlap}
\end{equation}

\begin{algorithm}[t]
\caption{\gls{tod} Admission Control and Routing (\gls{tovac})}\label{al:astar}

\begin{algorithmic}[1]
\Require  $ \{\nu_i\}_i, \{s_\nu\}_\nu,\{d_\nu\}_\nu,G$

\State $E(G)\setminus= \{e: \nu_e^{\max}(D,\mu,b,L)<1\}$\Comment{Prune roads} \label{al:prune}

\For{$\nu \in \{\nu_i\}_i $}
    \State $\mathcal{P}= A^*(G,s_\nu,d_\nu,\text{weight}=\omega_\nu,\text{heuristic=Haversine})$
    \State $X_{\mathcal{P}}^{\nu}=1$ if $\sum_{e\in \mathcal{P}} d(e)<\infty$ else $0$
\EndFor

\State \Return $\{X_{\mathcal{P}}^\nu\}_\nu$
\end{algorithmic}
\end{algorithm}

In conclusion, the algorithm will deny the remote operation of a vehicle if all the possible paths between the source and the destination include a road section without capacity. Otherwise, the vehicle that is requesting the route is admitted, as line 4 in Algorithm~\ref{al:astar} shows. 
It is worth remarking that, contrary to~\cite{routeplan2022},
\gls{tovac} considers the time dimension and does not
pitfall into static routing.

\section{Performance Evaluation}


For validation and insight into the behavior of our proposed admission control and routing algorithm, we have implemented \gls{tovac}
in Python using the NetworkX~\cite{Hagberg2008} package to
build Section~\ref{subsec:graph} capacity graph $G$.
Each road segment $e\in G$ corresponds to
an OpenStreetMap~\cite{osm} nodeway of Turin (Italy) city center.
Additionally, we resort to the tractable channel
model considered in~\cite{snc,coverage},
and use the \gls{sinr} \gls{ccdf}
expression obtained in Corollary~\ref{corol:snr}
to know the 99.999 \gls{sinr}.

For coverage information, we have used the open data provided by OpenCelliD\footnote{OpenCelliD, largest Open Database of Cell Towers \& Geolocation - by Unwired Labs, https://opencellid.org/.}, a database of cell towers. In particular,
we consider that 5G cell
coordinates correspond to
the location of LTE cell
towers of the network operator with MNC 1 in Turin.
If multiple road segments
$E'\subset E(G)$
are only served by a single
LTE cell
-- i.e. $\exists! r:
e\in E(r),\ \forall
e\in E'$ --
we consider
a 5G deployment with a
dedicated cell
per road.
We capture such
consideration by
setting 
 $\iota_r(e)=1, \forall e\in E'$.
Note the
consideration holds given
the necessity of denser
deployments in
cellular mmWave.


We have applied the NR configuration described in Table~\ref{tbl:nrsetup} to each cell.
In the experiments we consider
four possible channel 
bandwidths in the
\gls{fr} 2,
namely:
$B=[80, 160, 240,320]$\,MHz.
Note that, in Italy, the allocation to individual operators in the 26\,GHz band is 200\,MHz~\cite{italyfreq}, which means that
several of the considered
bandwidths are optimistic.

We conduct a comprehensive
set of experiments defined
as follows.
Each experiment
$\varepsilon$
consists
of
$V_\varepsilon=5v+1,\
v=0,\ldots,20$
concurrent vehicle requests, with
$S_\varepsilon=|\{s_i\}_{i=1}^{V_\varepsilon}|=\lceil V_\varepsilon/2^s\rceil,\ s\geq0$ sources;
and
$D_\varepsilon=|\{d_i\}_{i=1}^{V_\varepsilon}|=\lceil V_\varepsilon/2^d\rceil,\ d\geq0$ destinations.
Each vehicle $\nu_i$ is
randomly assigned to
a source $s_{\nu_i}\in\{s_i\}_{i=1}^{V_\varepsilon}$
and a destination $d_{\nu_i}\in
\{d_i\}_{i=1}^{V_\varepsilon}$.
Each experiment $\varepsilon$
restricts/allows all vehicles to
($i$) 
share same source and destination,
i.e. $2^d,2^s>V_\varepsilon$;
($ii$) have completely
different destinations
$d=s=0$; or
($iii$) have a mix
of shared/different
destinations
$0<d,s<V_\varepsilon$.


Only the uplink traffic, the most demanding, is analyzed, with each vehicle sending 25\,Mbps video stream traffic ($L=12$\,kb) with a \gls{pdb} of 5\,ms.
The speed of the vehicles is, in each
traversed road $e \in G$,
equal to the maximum speed of that road $e$.

\begin{table}[t]
    \caption{Experimental \gls{nr} setup}
    \label{tbl:nrsetup}
    \centering
    \resizebox{\columnwidth}{!}{
    \begin{tabular}{r l | r l}
        \toprule
        \textbf{Parameter} & \textbf{Value} & \textbf{Parameter} & \textbf{Value}\\
        \midrule
        Service bitrate $b$ & 25\,Mbps~\cite{3gppV2X} & Pkt. len. $L$ & 12\,kb\\
        \gls{pdb} & 5\,ms &  Numerology $\mu$ & 2~\cite{snc} \\
        $\text{NRB}_r(T_S^\mu,80\,MHz)$ & 108~RB~\cite{3gpp-nrb} & Freq. & FR2 26\,GHz~\cite{italy-assign}\\
        $OH$ & 0.14~\cite{3gpp-overhead} & Bandw. $B$ & $[80,320]$\,MHz~\cite{italy-assign}\\
        Pathloss exp. $\alpha$ & 4~\cite{mmwave-path-loss} & Decode $\epsilon_r$ & $10^{-5}$~\cite{finite-blocklength}\\
       
        \bottomrule
    \end{tabular}}
\end{table}
\begin{figure}[t]
    \begin{tikzpicture}
    \node[inner sep=0pt] (combi) at (0,0)
        {\includegraphics[width=\columnwidth]{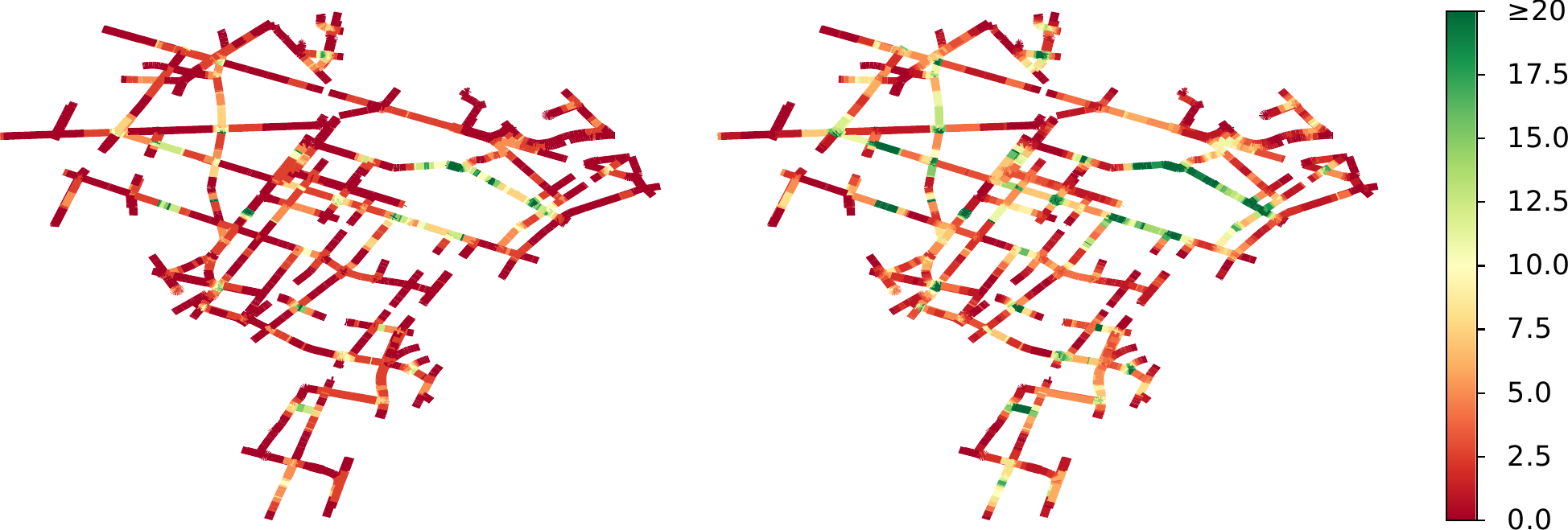}};
    \node[align=center] at ($(combi)+(-2.5,1.8)$) {\footnotesize 80\,MHz};
    \node[align=center] at ($(combi)+(1.75,1.8)$) {\footnotesize 320\,MHz};
    \node[align=center] at ($(combi)+(4.05,1.8)$) {\footnotesize $\nu_e^{\max}$};
    \end{tikzpicture}
    \vspace{-1.5em}
    \caption{Heatmaps with the maximum number
    of remote driving vehicles that can
    be operated with 99.999\,\% reliability
    in Turin. 
    }
    \label{fig:heatmap}
\end{figure}
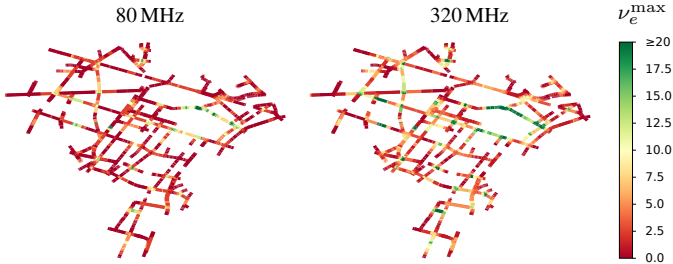

Fig.~\ref{fig:heatmap} shows two heatmaps of the area of Turin where we have evaluated \gls{tovac}. Each heatmap represents the number of vehicles that can be remotely driven on each road segment with 99.999\,\% reliability -- i.e., each heatmap represents the capacity graph $G$. Greener colors mean that more vehicles can be served in the corresponding road segment (the greenest color is dedicated to segments that allow $\nu_e^{\max}=20$ or more vehicles, with some cases reaching more than $\nu_e^{\max}=30$). In one heatmap, the available uplink bandwidth for the remote driving service is
$B=80$\,MHz, and in the other it is $B=320$\,MHz (the extreme cases). The figure shows the challenges of providing the
reliability and data rate required by the \gls{tod} service
if cells are co-located with LTE cell towers.

\begin{figure}[t]
    \input{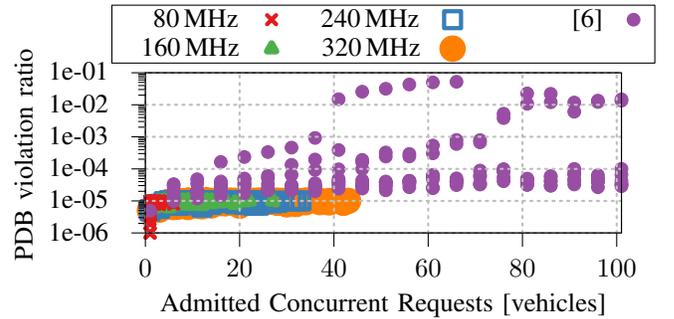}
    \vspace{-1em}
    \caption{\gls{pdb} violation ratio (y-axis) experienced among the
        admitted vehicles (x-axis). }
    \label{fig:pdbviolations}
\end{figure}
An important feature of \gls{tovac} is its ability to keep \gls{pdb} violations below the 99.999\,\% threshold.  This is illustrated in Fig.~\ref{fig:pdbviolations}. This figure shows, for the different
bandwidths $B$, the concurrent requests (\gls{tod} vehicles) admitted for each realization of the different experiments (defined by the number of vehicles requesting the \gls{tod} service), and the resulting \gls{pdb} violations. Using \gls{tovac}, the fraction of packets sent that result in PDB violations is always kept below $10^{-5}$. This is calculated by measuring the number of vehicles coinciding on the same road segment at the same time range, and the \gls{se} given the worst case \gls{sinr} and the resulting \gls{mcs} on the road segment. With \gls{tovac}, it never happens that more vehicles coincide on a road segment than those that guarantee the \gls{pdb}, and \gls{pdb} violations are the result of situations in which the \gls{sinr} does not achieve the expected values, thus, resulting in packet delays $D>5$\,ms. We also compare our approach with the solution in \cite{routeplan2022}, where, as shown in Fig.~\ref{fig:pdbviolations}, there are more vehicles admitted for tele-operation, in fact, all vehicle requests are admitted. However, the PDB violations are greater than $10^{-5}$ and can reach values close to $10^{-1}$ -- i.e., one out of ten \gls{tod} packets violate the \gls{pdb}. Note that in our implementation of~\cite{routeplan2022}, we do not reduce the speed of the vehicles.
Nevertheless, reducing the speed as required by~\cite{routeplan2022} would mean that the vehicles would be on the roads longer and more PDB violations would occur.

\begin{figure}[t]
    \input{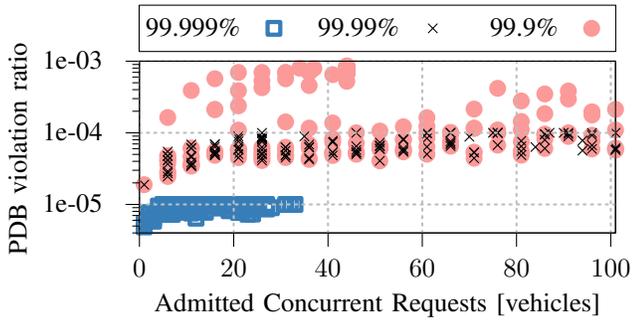}
    \vspace{-1em}
    \caption{\gls{pdb} violation ratio (y-axis) experienced among
    the admitted vehicles (x-axis) for different reliability requirements, with 240\,MHz bandwidth. }
    \label{fig:reliability}
\end{figure}

We also explored the effect of relaxing the 99.999\,\% reliability requirement. In Fig.~\ref{fig:reliability}, for the
$B=240$\,MHz bandwidth scenario, we compare the admitted vehicles for  99.999\,\%, 99.99\,\%, and 99.9\,\% reliability requirements. As it can be seen in the figure, the number of admitted vehicles for the different realizations of the experiments is significantly increased, both for 99.99\,\% and for 99.9\,\% reliability thresholds, in many cases admitting 100\,\% of the requests (compare Fig.~\ref{fig:reliability} with the case of the implementation of \cite{routeplan2022} in Fig.~\ref{fig:pdbviolations}). The downside is that the number of PDB violations also increases, although it is still smaller than the number of PDB violations in \cite{routeplan2022}.

Finally, it is interesting to highlight that even if the bandwidth required by the \gls{tod} service appears to be very high, the actual use of the radio resources is much smaller.
For example, the average consumption of radio resources
was of a 10.19\% in a stretch of Via Vittorio Emanuele II
-- a main road in Turin experiencing the highest
resource usage. Such result was observed
during the realization of experiment $\varepsilon$
with $V_\varepsilon=101$ vehicles
traveling from different sources and destinations ($s=d=0$)
and $B=240$\,MHz of available uplink bandwidth.
By employing an elastic slice~\cite{pi-road} to implement the reservation of resources for the \gls{tod} service, the unused resources could be exploited by other users while the vehicles are not using them, so providing the \gls{tod} service can be much less onerous for the operator than it might appear.


\section{Conclusion}
\noindent \gls{tod} services require reliable, high-bandwidth, and low-latency communications to ensure that the same safety level or better than traditional driving is achieved. In this letter, the \gls{tovac} algorithm has been proposed to prevent vehicular traffic from exhausting the available 5G network resources.
\gls{tovac} estimates the
99.999 \gls{sinr} percentile
that tele-operated vehicles
experiment at each road to
perform admission control
and routing decisions
that meet
the \gls{tod} \gls{qos}.
The obtained results show that all the vehicles admitted by \gls{tovac} obtain a service that meets the bandwidth, latency and reliability requirements defined in the \gls{3gpp} standards. Although it may seem that the number of vehicles admitted is very limited (around 50 in the best scenario), results show that
\gls{tovac} keeps a rather small network utilisation (up to about 10,19\% on average for the busiest road section).
This opens the door to keep pursuing the work on new and efficient online route planning algorithms, for example, to adapt the routes to the actual network conditions that vehicles are measuring along the route.


%





\ifCLASSOPTIONcaptionsoff
  \newpage
\fi



\newpage
\bibliographystyle{IEEEtran}
\bibliography{refs.bib}
%

%





\end{document}